\documentclass[aps]{revtex4-1}
\usepackage{geometry}                
\usepackage{graphicx}
\usepackage{amssymb}
\usepackage{epstopdf}
\begin{document}

\title{The electromagnetic decay of $^{250m}$No and the stability of neutron deficient Rf isotopes}

\author{A. Lopez-Martens, K. Hauschild,  R. Chakma\footnote{Present affiliation:ANL}}
\affiliation{IJCLab, CNRS/IN2P3 and Universit\'e Paris Saclay, Orsay, France }
\author{O. Dorvaux,  B. Gall,   K. Kessaci\footnote{Present affiliation:CEA}}
\affiliation{IPHC, CNRS/IN2P3 and Universit\'e de Strasbourg, Strasbourg, France}
\author{M.L. Chelnokov, V.I. Chepigin, A.V. Isaev, I.N. Izosimov, 
D.E. Katrasev,  A.A. Kuznetsova, O.N. Malyshev, A.G. Popeko, Yu.A. Popov, E.A. Sokol, A.I. Svirikhin, M. S. Tezekbayeva, A.V. Yeremin}
\affiliation{ FLNR, JINR, Dubna, Russia and Dubna State Univerity, Dubna, Russia}

\begin{abstract}
The electromagnetic decay of the $\approx$40 $\mu$s isomer of $^{250}$No has been investigated using the \textsc{Geant4} toolkit for the simulations of the interaction of particles through matter. It is concluded that the decay does not follow the pattern established in the lighter isotones, where the isomer decays directly to members of the ground state rotational band. An alternative scenario is proposed. The implications on the location of the isotopic border for neutron deficient Rf isotopes are discussed.

\end{abstract}
\maketitle

\section{Introduction}

Very heavy and super heavy nuclei owe their stability against fission to quantum shell effects, which dig one or more pockets in the potential energy surface of the nucleus at a given deformation thereby providing a barrier or multiple barriers against spontaneous fission. Other effects come into play, such as the conservation of quantum numbers (specialization energy) and superfluidity of the system in the fission process. This is why the fission half lives of ground states in nuclei with an odd number of protons (Z) or an odd number of neutrons (N) are known to be much longer than the ones of their even-even neighbours.  This effect is characterised by a fission hindrance factor ($HF$), which is defined as the ratio of the fission half life in the odd-Z or odd-N system, $T_{SF,gs}(Z,N)$, divided by the geometric mean of the fission half lives of the two neighbouring even-even nuclei; $(T_{SF,gs}(Z,N-1) \times T_{SF,gs}(Z,N+1))^{1/2}$ for odd-N nuclei or $(T_{SF,gs}(Z-1,N) \times T_{SF,gs}(Z+1,N))^{1/2}$ for odd-Z nuclei.  $HF$s are typically of the order of 10$^2$-10$^5$, depending on the particular microscopic configuration of the odd proton or neutron. However, as evidenced by the available data and stated in \cite{Hessberger17}, there is no connection between the magnitude of the spin of the odd nucleon and the measured $HF$. 

Reduced pairing and conservation of angular momentum on the pathway to fission is also the reason why multi-quasi-particle states, such as $K$ isomers, are thought to be more stable against fission than the ground state (where $K$ is the projection of the total angular momentum onto the symmetry axis of the nucleus). Populating these states should therefore lead to an enhanced survival probability of super heavy nuclei \cite{Xu,Jachimowicz,Clark}. Until recently there was no obvious experimental evidence that high-$K$ states had longer fission times than the corresponding ground state, except for $^{250}$No (see figure 13 of ref. \cite{Kondev}). However, new data for $^{254}$Rf \cite{David}, $^{250}$No \cite{Kallunkathariyil,Batar250No} and more recently $^{252}$Rf \cite{Batar252Rf} have shown that the fission hindrances (defined in this case as $HF=T_{SF,iso}/T_{SF,gs}$) of high-$K$ states in these nuclei are considerably larger than one.

The influence of shell effects on fission half lives is clearly observed in the plot of fission half-lives of even-even very heavy and super heavy nuclei as a function of neutron number N (see figure 10 of ref. \cite{Hessberger17}), where fission rates are seen to increase on either side of N=152 for the heaviest systems. This is especially visible for the No isotopic chain, where the lifetime drops by 10 orders of magnitude from $^{254}$No to $^{250}$No. The trend in fission lifetimes in going from $^{256}$Rf to $^{254}$Rf follows the one of the No isotones, accelerated by an average factor of the order of 5~10$^6$ \cite{LopezRf}. In $^{253}$Rf, the most neutron deficient known odd-A Rf isotope, recent experimental data has shown the existence of two nearly degenerate fissioning states \cite{Batar253Rf,LopezRf}: the ground state and a very low-lying isomeric state. On the basis of the assignment of a spin 7/2 instead of 1/2 to the faster fissioning state observed in $^{253}$Rf, and given the $HF$ of the same 7/2 state measured in $^{251}$No, a sub-ps fission half life was foreseen for the ground state of $^{252}$Rf, putting it out of reach of current experimental capabilities \cite{LopezRf}. 

The spin assignment in $^{253}$Rf led to much debate \cite{Rogov1} as it is generally assumed that higher-spin states survive better against fission than lower-spins states. Indeed, within the dinuclear system model \cite{Rogov2}, states with a large spin live longer than states with a smaller spin due to the inherent larger centrifugal barrier. In ref. \cite{Moller} the spin dependence is more nuanced as the effect is related to the distance to the Fermi surface during fission of the orbital occupied by the odd particle: ``Because there are generally fewer orbitals with high angular momentum than with low angular momentum, the orbital for an odd particle with high angular momentum will be farther away from the Fermi surface than that for a particle with low angular momentum''.  How the spin of the odd particle will affect the fission time therefore depends on the specific details of the Nilsson diagram on the pathway to fission. Moreover, as shown in ref. \cite{Rodriguez}, where the fission properties of different low-lying spin states in odd No isotopes have been investigated with constrained mean-field calculations based on the Gogny interaction, the pairing quenching induced by the odd particle also depends on the properties of the considered one-quasiparticle configuration. It is to be noted that though the calculated fission lifetimes in ref. \cite{Rodriguez} vary substantially with the model parameters, no systematic larger fission half lives are obtained for the higher spin states compared to the lower-spin states. 

The spin-assignment controversy in $^{253}$Rf and the consequent sub-ps lifetime estimate for $^{252}$Rf seemingly ended with the discovery paper of $^{252}$Rf \cite{Batar252Rf}. From analogy to the decay of a 6$^+$ isomer in the lighter isotones $^{242}$Pu \cite{Pu}, $^{244}$Cm \cite{Cm} and $^{250}$No \cite{Tezekbayeva}, the activities observed in $^{252}$Rf were assigned to a 13 $\mu$s 6$^+$ isomer with a fission branch $\le$90$\%$ and to a ground state with a 60 ns fission half life, in line with the theoretical predictions of Smolanczuk et al. \cite{Smolanczuk}. However, the controversial spin assignments in $^{253}$Rf were later confirmed in the discovery paper of $^{257}$Sg \cite{Batar257Sg} since the properties of the $\alpha$ decay of $^{257}$Sg to $^{253}$Rf support the high-spin assignment to the faster fissioning state in $^{253}$Rf. From the spin assignment of the ground state of the newly-discovered $^{257}$Sg isotope, the authors also inferred a sudden fall in the fission half life of the hitherto unknown $^{256}$Sg isotope by 3 orders of magnitude compared to the theoretical predictions of ref. \cite{Smolanczuk}. Unfortunately, the impact of the confirmed spin assignments in $^{253}$Rf on the interpretation of the $^{252}$Rf data was not discussed in ref. \cite{Batar257Sg}.  

In this paper we present a detailed analysis of the electromagnetic decay properties of the $\approx$40 $\mu$s supposed 6$^+$ isomer of $^{250}$No \cite{Belozerov,Peterson,Svirikhin,Kallunkathariyil,Batar250No,Tezekbayeva}. The \textsc{Geant4} simulation framework \cite{geant4} was used to model the isomeric emission of photons, internal-conversion electrons, Auger electrons and X rays as well as their interactions in the sensitive volumes of the GABRIELA detector array \cite{Hauschild,Chakma,LopezGabriela}. The well-established electromagnetic decay of the 8$^-$ isomer in $^{252}$No \cite{Sulignano,Sulignano2}, produced in the same experiment, was used to benchmark the simulations. 

\section{Experimental details}

\begin{figure}[htbp]
   \centering
  \includegraphics [angle=0,width=0.5\textwidth]{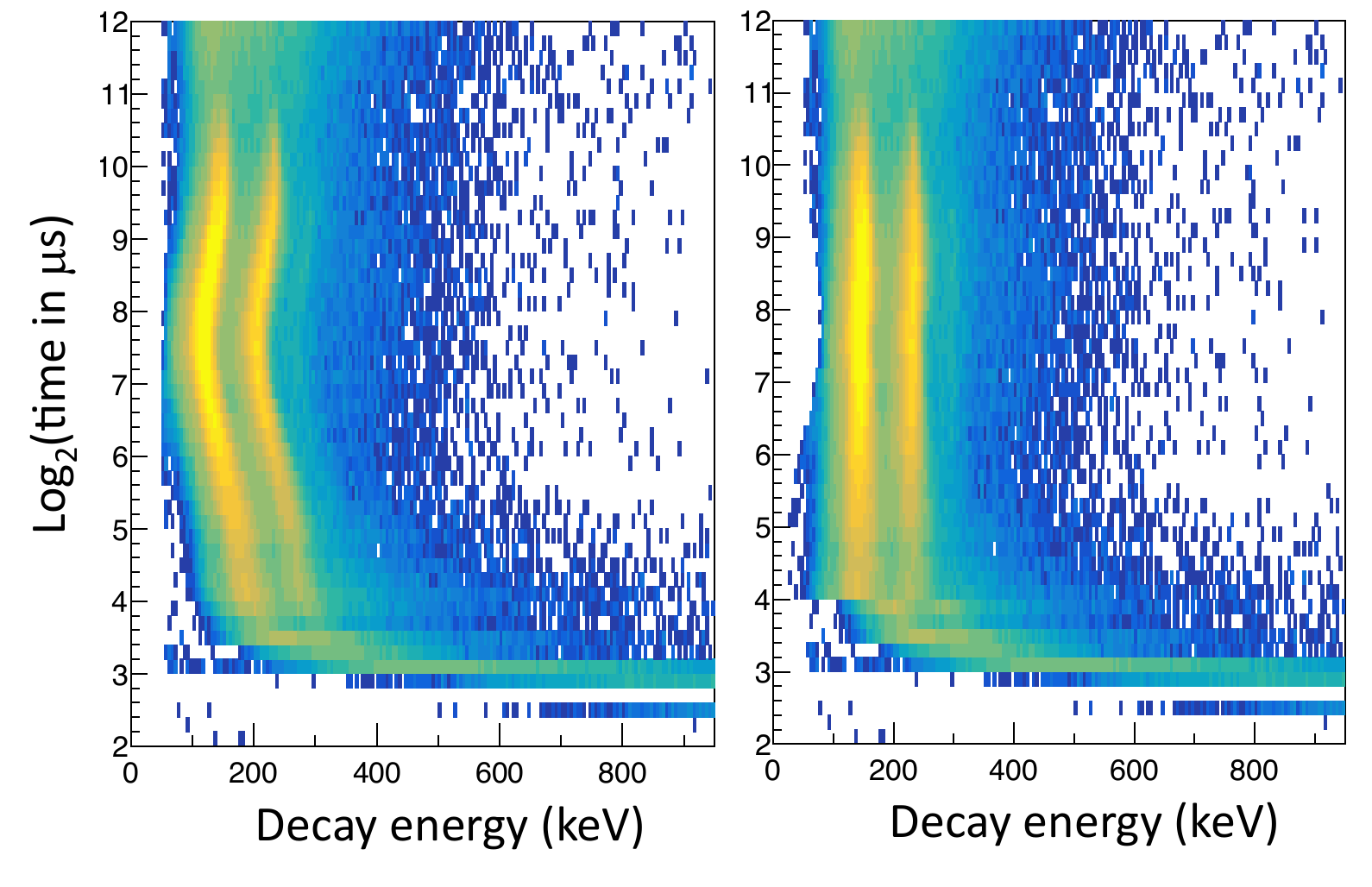} 
   \caption{Time difference (in $\mu$s) between the implantation signal of a $^{207}$Rn nucleus and its subsequent decay as a function of the measured decay energy. The two structures observed correspond to the K and LMN+ conversion of the 234~keV M2 transition deexciting the know 13/2$^+$ isomer in $^{207}$Rn \cite{Rezanka}. The right figure has been corrected for the pileup of the recoil and decay signals down to 16 $\mu$s (log$_{2}$(time)=4)}
   \label{fig0}
\end{figure}

The experiment was carried out in early 2019 and was the object of a preliminary report published in ref. \cite{Tezekbayeva}.
To produce $^{250}$No nuclei, the reaction $^{204}$Pb($^{48}$Ca,2n) was used at an average cyclotron beam energy of 224-225 MeV. Ti-backed targets of $^{204}$PbS with 99.94$\%$ $^{204}$Pb enrichment and average thickness 470(50) $\mu$g/cm$^{2}$ were used. The targets were rotated to withstand the intense flux of $^{48}$Ca ions provided by the U400 cyclotron at FLNR, JINR. A total dose of 2.6 10$^{18}$ impinging particles was recorded during the time of the experiment. For the production of $^{252}$No nuclei, $^{206}$PbS targets with thickness 406(40) $\mu$g/cm$^2$ and an isotopic enrichment of 99.5$\%$ were utilised. The beam dose accumulated on the $^{206}$Pb targets was 4.6 10$^{17}$. 
The heavy nuclei produced in the fusion-evaporation reactions were separated by the recoil separator SHELS \cite{Popeko}. They then passed through the 2 emissive foils of a Time of Flight detector and through a degrader foil and were implanted into the 10$\times$10 cm$^{2}$ Double-sided Silicon Strip Detector (DSSD) of an upgraded version of the GABRIELA detector array \cite{Chakma,LopezGabriela}. Surrounding the DSSD, in the upstream direction, 8 smaller-sized DSSDs forming a tunnel were used to detect escaping $\alpha$ particles and fission fragments as well as internal conversion electrons (ICEs). $\gamma$ and X rays were detected with a ring of 4 coaxial Ge detectors and a large clover detector placed just behind the DSSD. All Ge detectors were surrounded by a BGO shield, which provides a flag in case of coincident Ge and BGO events. 
\begin{figure}[htbp]
   \centering
   \includegraphics [angle=0,width=0.5\textwidth]{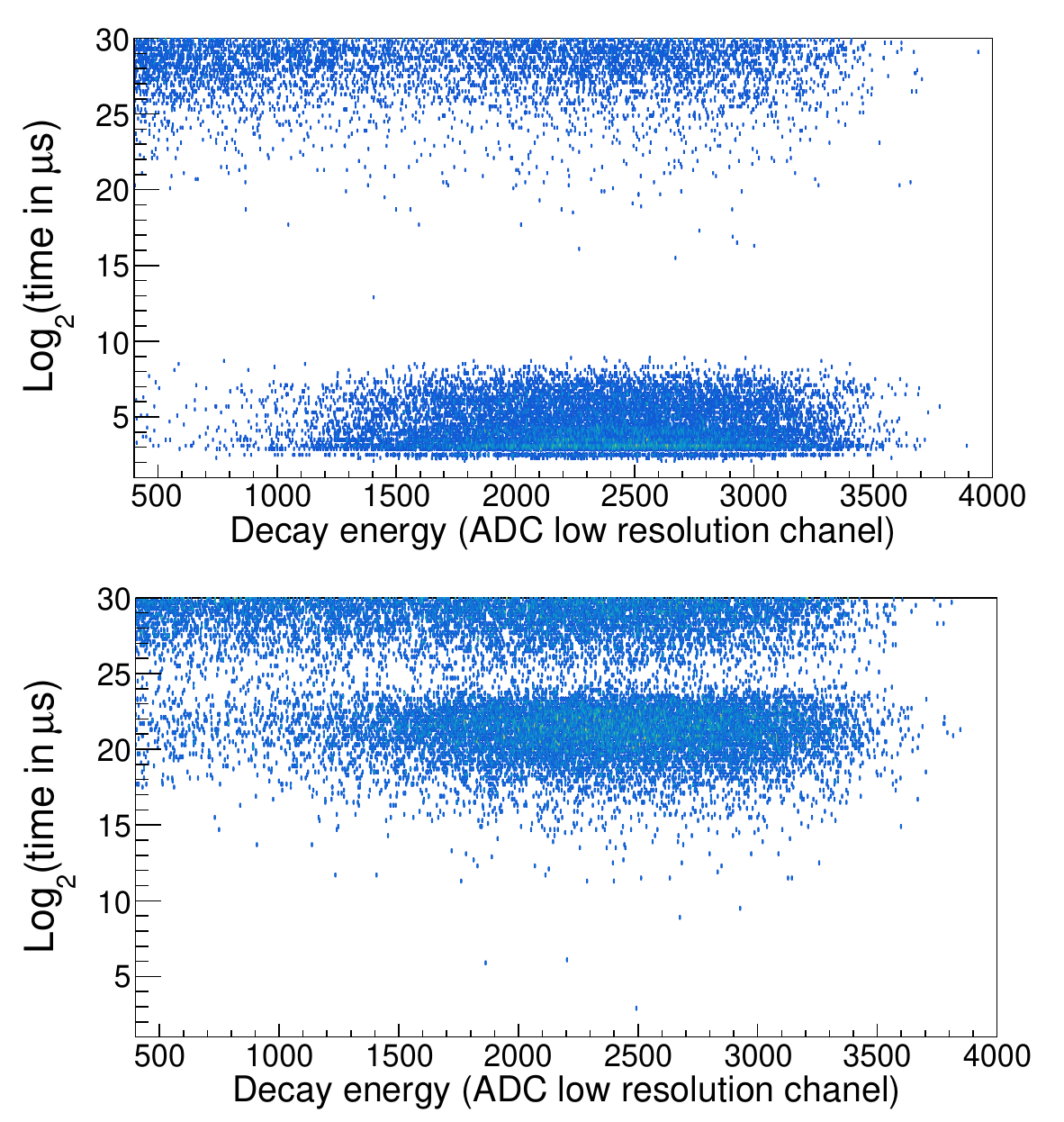} 
   \caption{Time difference (in $\mu$s) between the implantation of a recoil and the detection of a subsequent decay in the same pixel of the DSSD as a function of the back strip energy (in low-resolution ADC channels with a $\sim$250 MeV range) in the case of top panel) the $^{204}$Pb($^{48}$Ca,2n)$^{250}$No reaction and bottom panel, the $^{206}$Pb($^{48}$Ca,2n)$^{252}$No reaction.  }
   \label{fig2}
\end{figure}
The signals from all the detectors were time-stamped with a 1$~\mu$s clock. The amplification range of the analog electronics was set to detect ICEs and $\alpha$ particles up to 25 MeV on the front side of the DSSD and $\alpha$ particles and fission events up to 250 MeV in the back strips. This particular configuration leads to quite high thresholds ($\sim$250 keV) for reconstructed pixel events (called [FB]), which require the coincidence between front and back strips. The DSSD events where only a signal on the front strip is recorded are called [F] and exhibit much lower thresholds ($\sim$80-100 keV). With its current instrumentation the dead time between recorded events in the DSSD lies between 4 and 8 $\mu$s (as visible in figures \ref{fig2} and \ref{fig3}).
\begin{figure}[th]
   \centering
   \includegraphics [angle=0,width=0.3\textwidth]{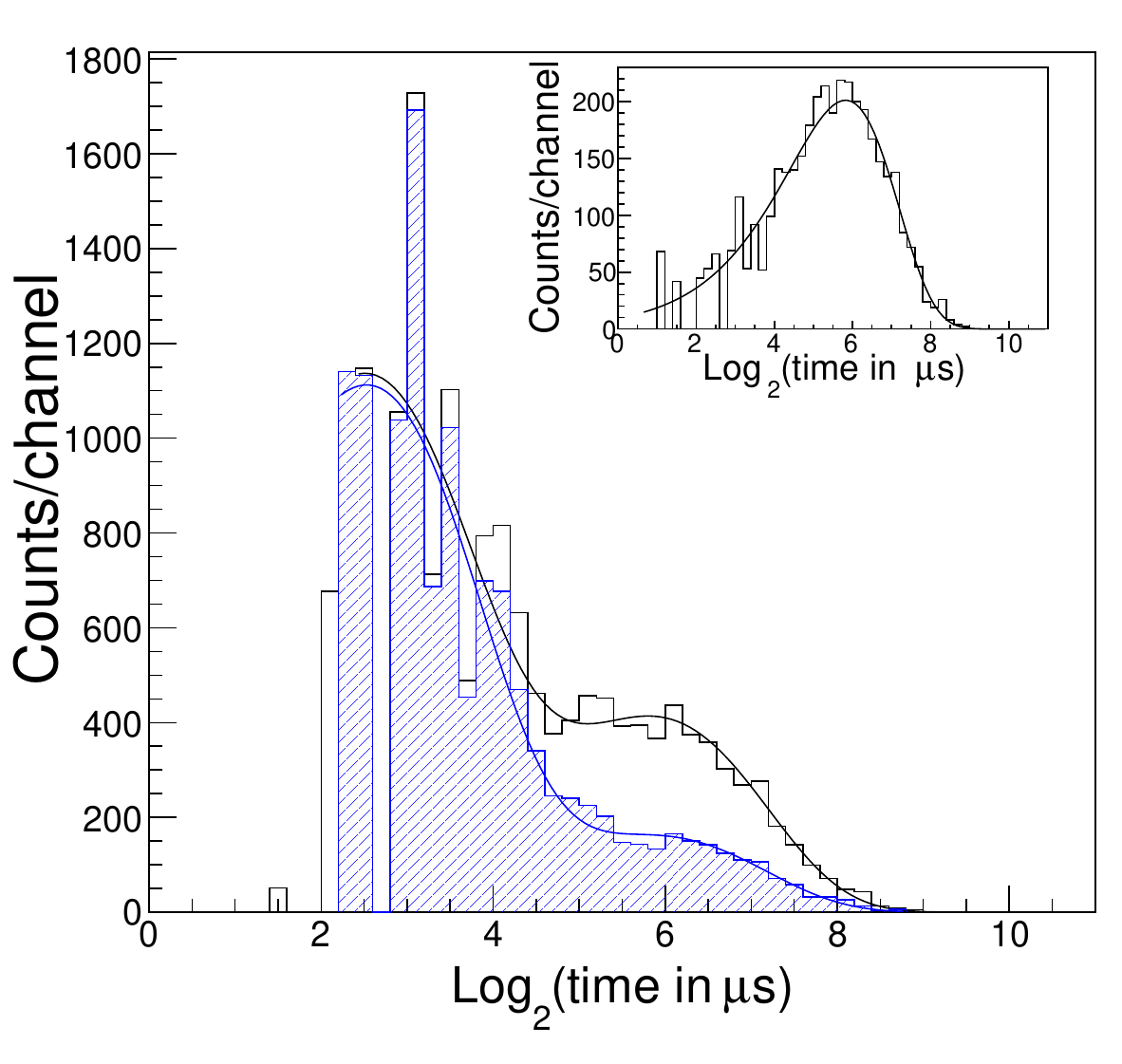} 
   \caption{The distribution of the time difference (in $\mu$s) between the implantation of a $^{250}$No nucleus and the subsequent detection of a fission signal in the same pixel of the DSSD is shown in black. The blue histogram corresponds to the time distribution of the fission events for which no front-strip event [F] or electromagnetic emission in the Ge or tunnel detectors is detected in between the recoil implantation and the fission signal. The time distribution of intermediate [F], Ge or tunnel events is shown in the inset. See the text for details.}
   \label{fig3}
\end{figure}

Energy and absolute efficiency calibration runs for ICEs, $\gamma$ rays and $\alpha$ particles were performed with $^{48}$Ca-induced reactions on stationary $^{164}$Dy and $^{174,176}$Yb targets leading to the production of $^{207m}$Rn and $\alpha$-emitting Th and Ra isotopes. The pileup energy correction of decay events following a recoil-implantation signal could be established down to an implantation-decay time difference of 16$~\mu$s using the internal-conversion emission from the 13/2$^+$ isomer of $^{207}$Rn \cite{Rezanka} (see figure \ref{fig0}). This limit was set on all the decay events detected after a recoil implantation for the energy measurements presented in this work. 
\begin{figure}[htbp]
   \centering
   \includegraphics [angle=0,width=0.3\textwidth]{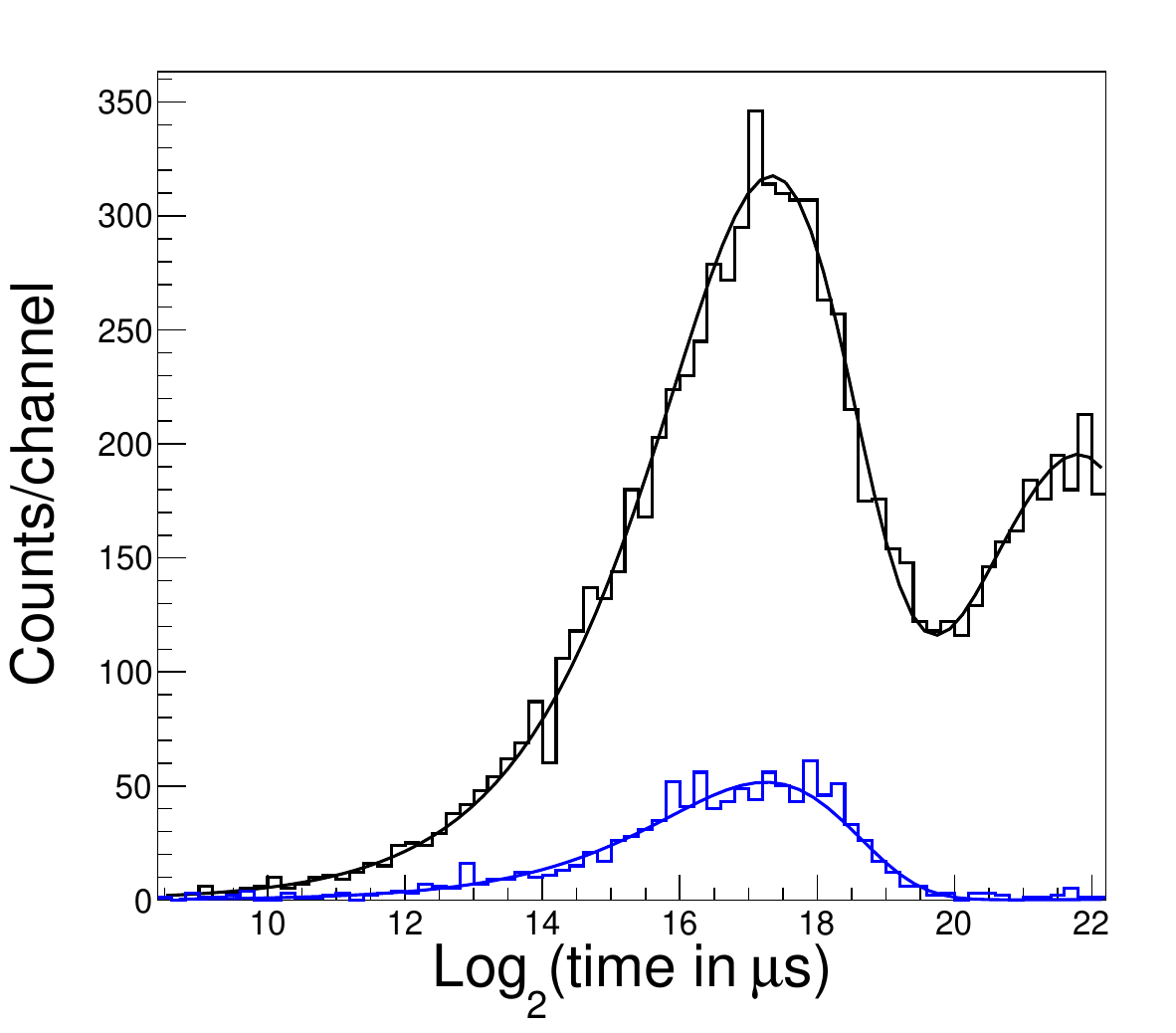} 
   \caption{Distribution of the time difference (in $\mu$s) between the implantation of a $^{252}$No nucleus and the subsequent detection of a [F] (black) or [FB] (blue) signal preceding the $\alpha$ decay or fission of the ground state and corresponding fits. }
   \label{fig4}
\end{figure}

\section{Results}
\begin{figure*}[th]
   \centering
  \includegraphics [angle=0,width=0.9\textwidth]{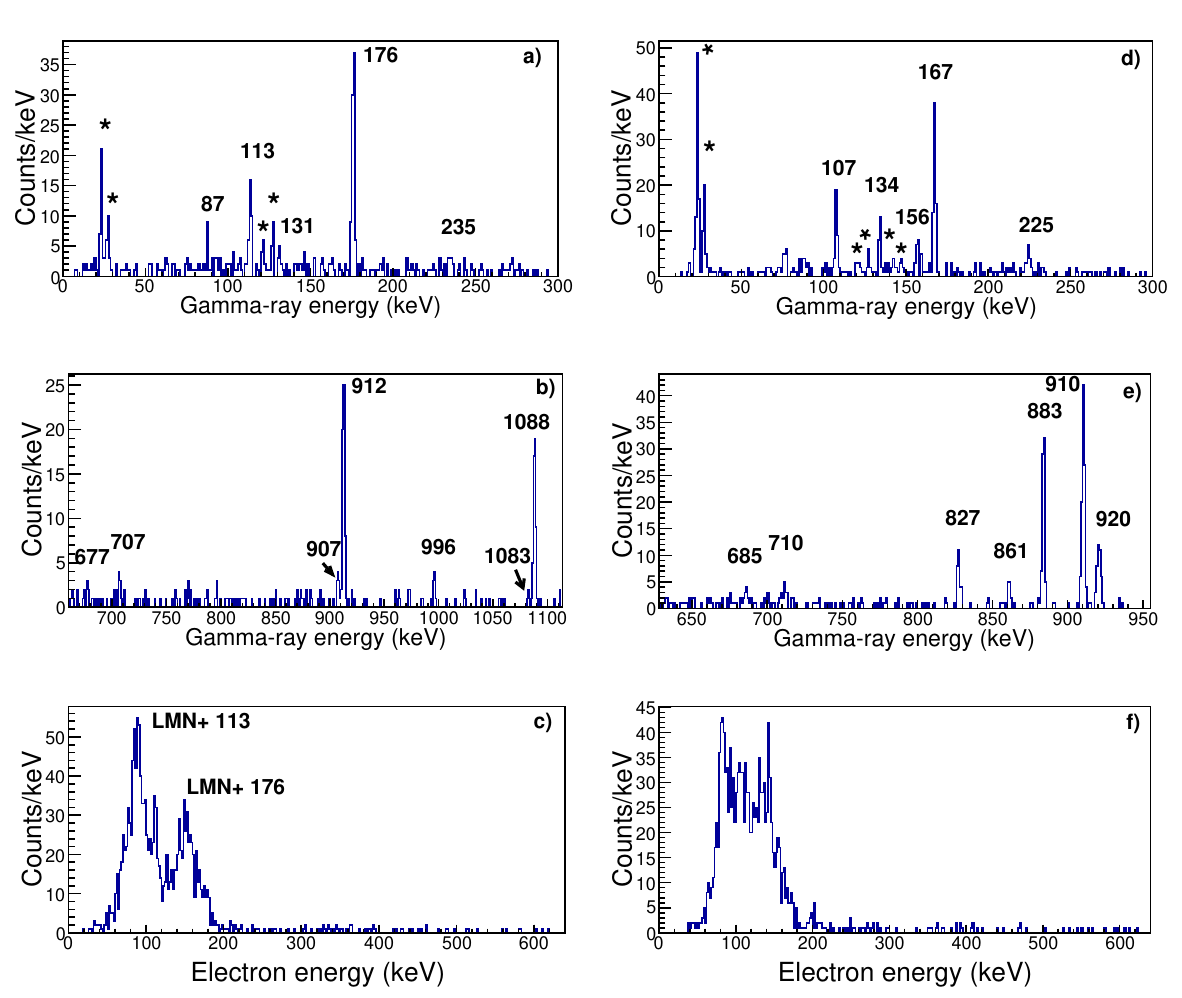} 
   \caption{a-b) Spectrum of $\gamma$ rays and c) ICEs emitted in the decay of $^{250m}$No. Panels d-f) show the same spectra in the case of the decay of $^{252m}$No. Stars indicate No L and K X rays. }
   \label{fig5}
\end{figure*}
The recoil-position-correlated fission events observed in the $^{250}$No and $^{252}$No runs are displayed in figure \ref{fig2}. The large signals associated to the fission of the ground state of $^{252}$No are clearly visible and well separated from randomly correlated fission events.  A half-life of 2.38(2) s was measured, in agreement with the evaluated value \cite{NNDC}. In contrast, fission events observed during the production of $^{250}$No occur within less than $log_2(\Delta t)=8$, i.e within less than 256 $\mu$s of the recoil implantation signal. Their time distribution is given in figure \ref{fig3}. Fits to the distribution yield 2 half-lives: 3.9(1)$~\mu$s for the faster decay component and 39.1(6)$~\mu$s for the slower one. These values are in good agreement with the literature values of the half lives of $^{250}$No and $^{250m}$No \cite{NNDC}. 

Though the slower fission events in figure \ref{fig3} are not preceded by the detection of a low-energy [FB] event in the DSSD, a significant fraction of them are preceded by a low-energy [F] signal in the same front strip of the DSSD or by the detection of a $\gamma$ ray in the Ge array and/or an ICE in the tunnel detectors with no accompanying [F] signal. The time distribution of these intermediate events is shown in the inset of figure \ref{fig3} and yields a half-life of 41.9(9) $\mu$s confirming the fact that they correspond to the electromagnetic decay of $^{250m}$No. In the case of $^{252}$No, the time distributions of isomeric [FB] and [F] events preceding the ground-state $\alpha$ decay or fission are shown in figure \ref{fig4}. The fits to the time distributions yield an average half life of 113(2) ms, in line with the evaluated half life of $^{252m}$No \cite{NNDC}.

The left panels of figure \ref{fig5} show the spectra of $\gamma$ rays and ICEs emitted by $^{250}$No and detected either alone after the recoil implantation or in coincidence with an isomeric [F] or [FB] signal in the DSSD. Four peaks clearly dominate the $\gamma$-ray spectrum at 113, 176, 912 and 1088 keV. At low energy, No LX rays appear below 40 keV and KX rays at 121 and 127 keV. At least 7 other much weaker lines are observed at 87 and 235 keV and at 677, 707, 907, 996 and 1083 keV. There is also potentially a 131 keV transition and it should be noted that the ratio of K$_{\alpha 1}$ to K$_{\alpha 2}$  intensities suggests that the 121 keV may be a doublet.

\begin{figure}[htbp]
   \centering
  \includegraphics [angle=0,width=0.5\textwidth]{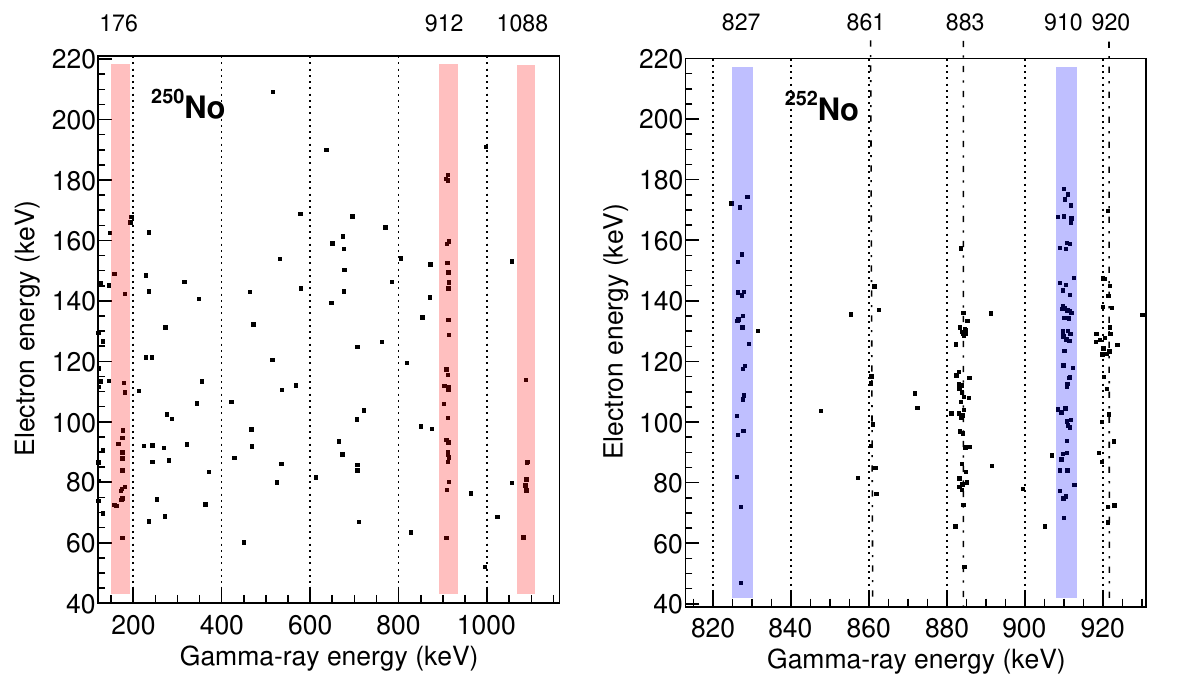} 
   \caption{Electron energies vs. coincident $\gamma$-ray energies observed in the decay of $^{250m}$No and $^{252m}$No. See the text for details on the highlighted coincidences. }
   \label{fig7}
\end{figure}

In panel c) of figure \ref{fig5}, two structures at energies corresponding to the LMN$^+$ internal conversion of the 113 and 176$~$keV transition, are clearly observed. The lowest peak lies at an energy for which the efficiency of the tunnel detectors is already affected by the thresholds. The intensity of the other peak on the other hand yields internal conversion coefficients $\alpha_{LMN+}$=3.7(9) and $\alpha_{L}$=2.3(7) for the 176 keV transition. These values are compatible with an E2 nature of the 176 keV transition ($\alpha_{LMN+}$=2.49 and $\alpha_{L}$=1.78 \cite{BRICC}).
The $\gamma$-ICE coincidence matrix of figure \ref{fig7} shows that the 1088 keV $\gamma$-ray line is not in coincidence with the LMN$^+$ conversion of the 176 keV transition while the 912 keV line clearly is and that the 176 keV $\gamma$ ray is coincident with LMN+ ICEs of the 113 keV transition. 

Given the $\approx$0.1 s half life of the isomer in $^{252}$No and the average Ge detector and tunnel DSSD rates, it was not possible, as for $^{250}$No, to isolate the spectrum of isomeric $\gamma$ and ICE decays detected alone after the recoil implant. Only the spectra of photons and ICEs detected in coincidence with [F] and [FB] isomeric events of $^{252m}$No could be cleanly produced. These are shown in the right panels of figure \ref{fig5}.  All the $\gamma$ rays observed by Sulignano et al. \cite{Sulignano} are clearly visible (see the top decay scheme of figure \ref{schemes}). The spectrum of electron energies in panel F is not resolvable given the experimental resolution and the many transitions from the ground-state band as well as from the excited rotational band that contribute to it. However, the $\gamma$-ICE coincidence matrix of figure \ref{fig7} supports the established decay scheme as the 827 and 910 keV transitions are observed in coincidence with the LMN+ ICEs of the 167~keV 6$^+-4^+$ ground state band transition, while the energies of the ICEs in coincidence with the other high energy lines do not extend beyond $\sim$150 keV.
\section{Discussion}
\begin{figure}[htbp]
   \centering
   \includegraphics [angle=0,width=0.5\textwidth]{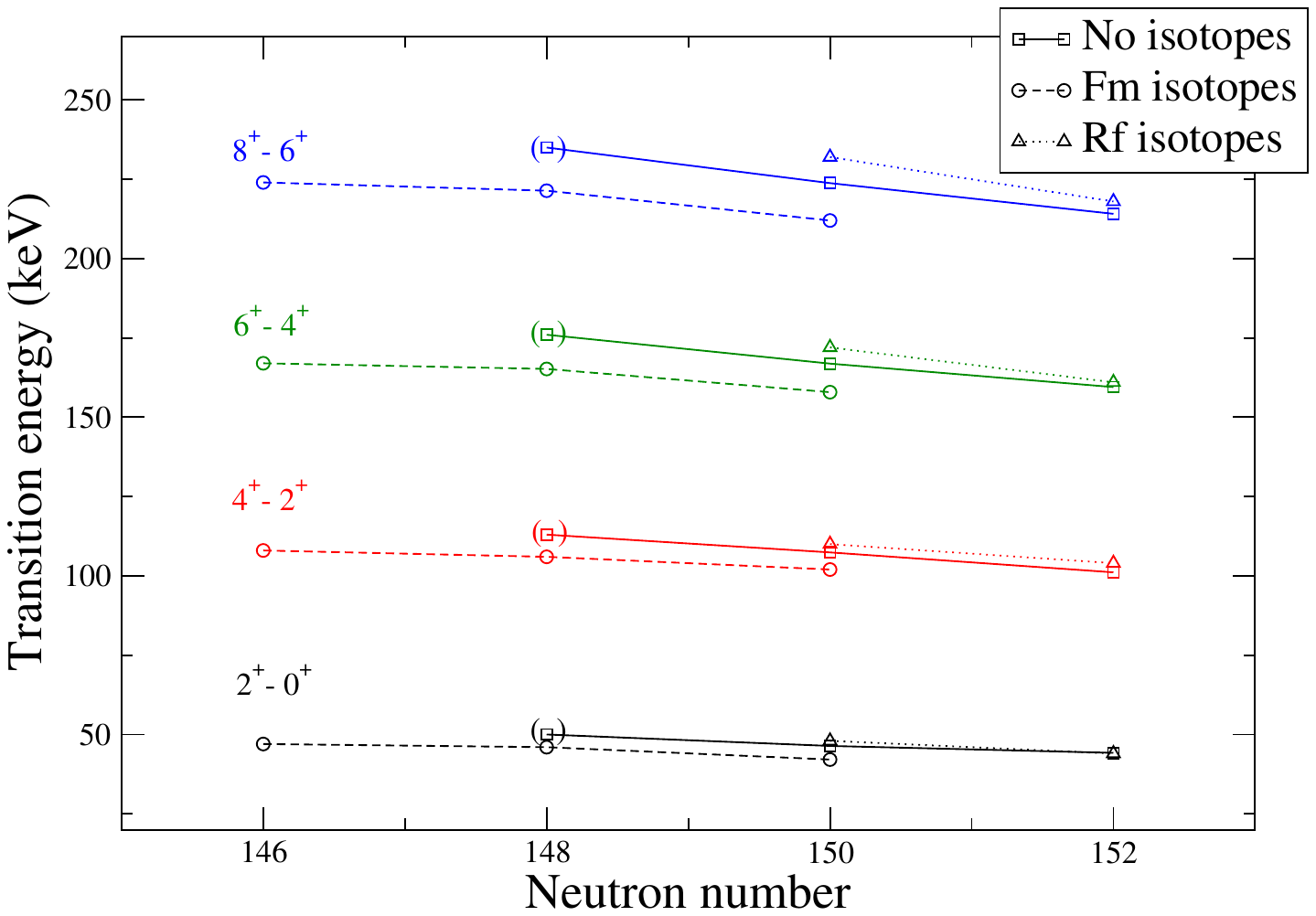} 
   \caption{Experimentally-deduced transition energies connecting members of the ground state rotational band of even-even No \cite{Reiter,Leino,Butler,Eeckhaudt,Herzberg}, Fm \cite{Greenlees,Ketelhut,Piot} and Rf \cite{GreenleesRf,Seweryniak} isotopes as a function of neutron number. The suggested energies for the ground state rotational band in $^{250}$No are marked with parentheses.}
   \label{bands}
\end{figure}

Given the trend in ground-state band transitions as one moves away from the N=152 shell closure (see figure \ref{bands}), the 235, 176 and 113 keV transitions are assigned to the 8$^{+}-6^+$ ,  6$^+-4^+$ and 4$^+-2^+$ transitions of the ground state rotational band of $^{250}$No. Based on these assumptions, a $\sim$50 keV energy for the 2$^+-0^+$ transition can be deduced using a Harris fit \cite{Harris}. 

In the lighter N=148 isotones $^{242}$Pu \cite{Pu}, and $^{244}$Cm \cite{Cm}, a low-lying 6$^+$ isomer, interpreted as a two quasi-neutron excitation involving the 5/2$^+$[622] and 7/2$^+$[624] Nilsson states, decays directly to the 6$^+$ and 4$^+$ states of the ground-state rotational band, and also to the 8$^+$ state in the case of $^{244}$Cm. This state is predicted to be the lowest high-$K$ state also in $^{250}$No, lying well below the two quasi-neutron 8$^-$ and 7$^-$ states \cite{Liu}. Given the difference in energy between the 912 and 1088 keV transitions and between the 912 and 678 keV transitions and the observed $\gamma$-ICE coincidences, it is logical to assume that the isomer in $^{250}$No is of the same nature and decays to the 6$^+$ member of the ground-state rotational band by emitting a 912 keV transition, to the 4$^+$ state via the 1088 keV transition and that it decays weakly to the 8$^+$ state through the 677 keV transition (see the bottom decay scheme of figure \ref{schemes}). 

To validate the proposed decay scheme, \textsc{Geant4} \cite{geant4} simulations of the electromagnetic decay of $^{250m}$No were carried out. In order to verify the threshold modelling for the DSSD detectors, extract the evaporation-residue implantation depth and benchmark the method, simulations of the decay of $^{252m}$No were also performed. The simulations include a realistic description of the active volumes of the detectors and materials surrounding them as well as the effects of the electronics multiplexing, which have been validated in ref. \cite{Chakma}. The effects due to the electronics thresholds of the different DSSDs of GABRIELA are modelled with a sigmoid function. The simulations include all the radiative and non-radiative processes accompanying internal conversion \cite{Chakma25}. 
\begin{figure}[htbp]
   \centering
  \includegraphics [angle=0,width=0.4\textwidth]{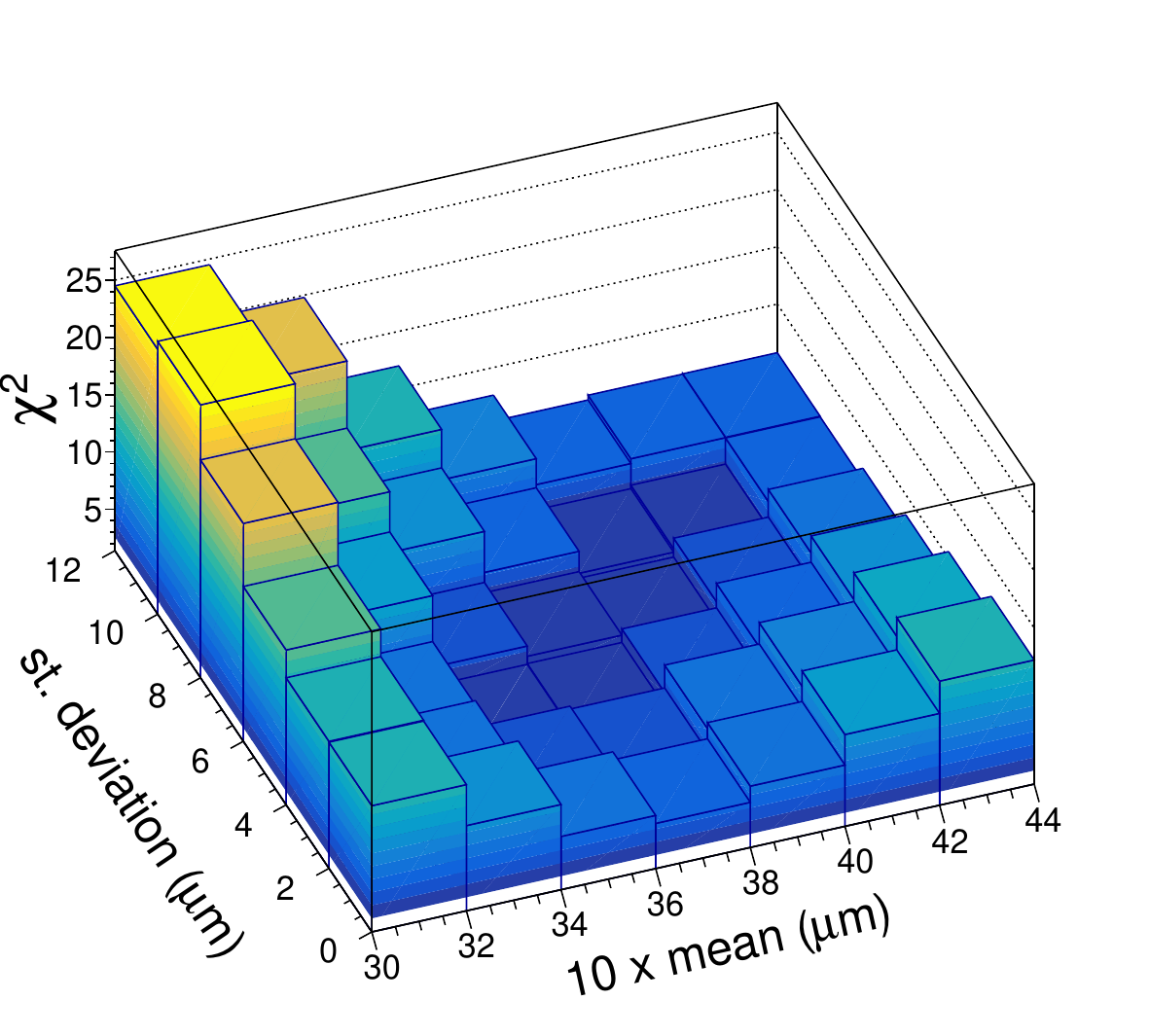} 
   \caption{$\chi^2$ difference between the experimental and simulated energy spectra of $^{252}$No escaping $\alpha$ particles as a function of the mean and the standard deviation of a Gaussian implantation depth profile.}
   \label{fig8}
\end{figure}
The experimental horizontal and vertical distributions of implanted heavy nuclei in the implantation DSSD were used as the (x,y) distribution of the source of radiation. The implantation depth of the heavy ions (z position) was determined according to the procedure described and illustrated in refs. \cite{Chakma,LopezGabriela} by adjusting the average and standard deviation of a gaussian implantation profile so as to reproduce the energy depositions of the $^{252}$No $\alpha$ particles, which escape from the DSSD. Figure \ref{fig8} shows the $\chi^2$ difference between the experimental and simulated energy spectra of $^{252}$No escaping $\alpha$ particles as a function of the mean and the standard deviation of a Gaussian implantation depth profile. A minimum is observed at a mean implantation depth and standard deviation of 3.7 $\mu$m and 0.7 $\mu$m respectively.
\begin{figure}[htbp]
   \centering
  \includegraphics [angle=0,width=0.5\textwidth]{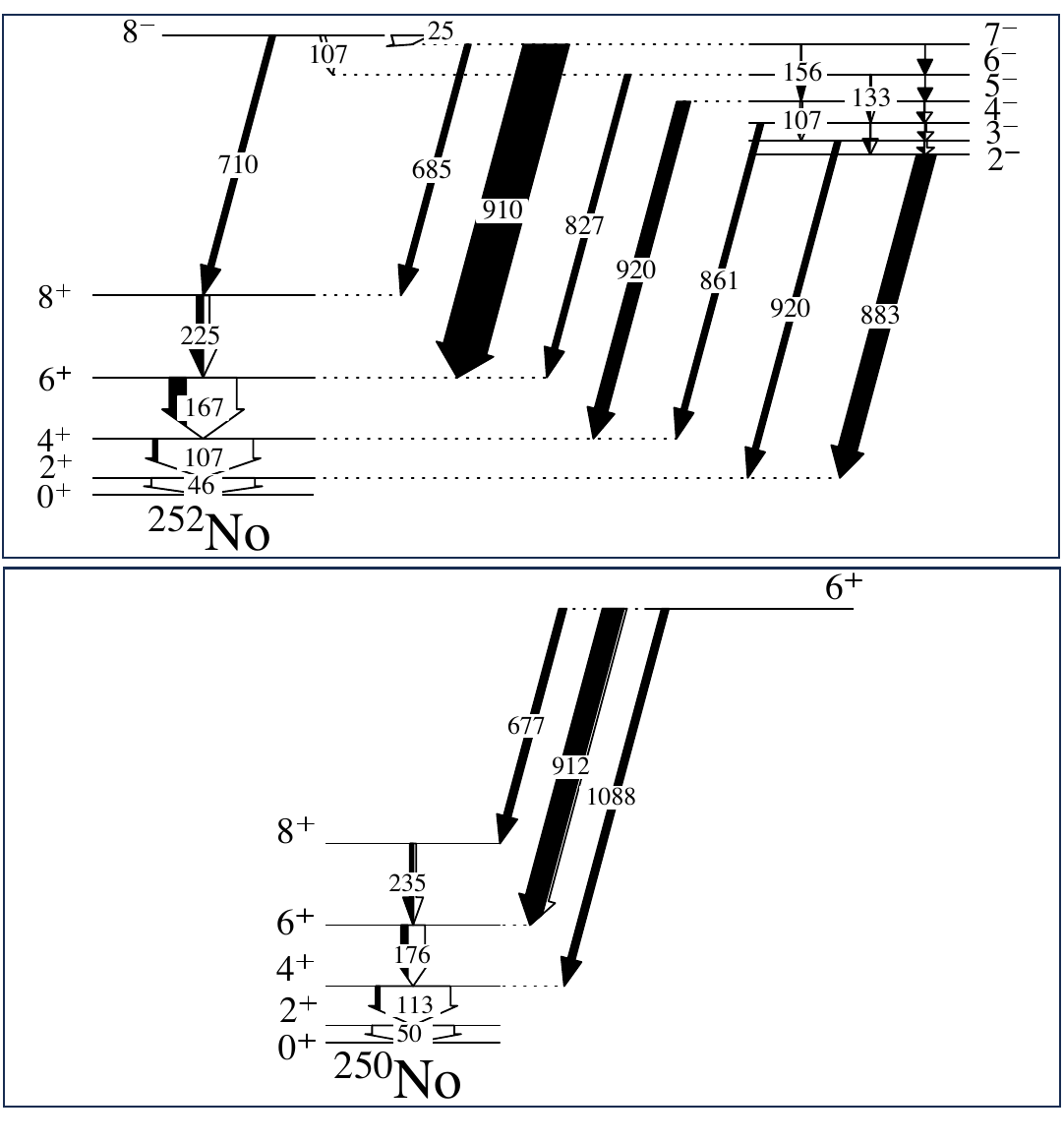} 
  
   \caption{Decay schemes of $^{252m}$No established in ref. \cite{Sulignano} and of $^{250m}$No updated from ref. \cite{Tezekbayeva}. The fraction of white filling in the arrows reflects the amount of internal conversion. For clarity, the thicknesses of the arrows of the transitions to and within the 2$^-$ band have not been scaled to the transition intensities.}
   \label{schemes}
\end{figure}

The decay schemes used to perform the simulations are shown in figure \ref{schemes}. The properties of the electromagnetic decay within the 2$^-$ rotational band of $^{252}$No have been calculated using a typical axial deformation parameter $\beta_2$=0.25 \cite{Raeder} and the gyromagnetic factor established for the 2$^-$ band in $^{250}$Fm \cite{Greenlees}. These parameters define the M1+E2 and E2 transition rates.

\begin{figure}[htbp]
   \centering
  \includegraphics [angle=0,width=0.5\textwidth]{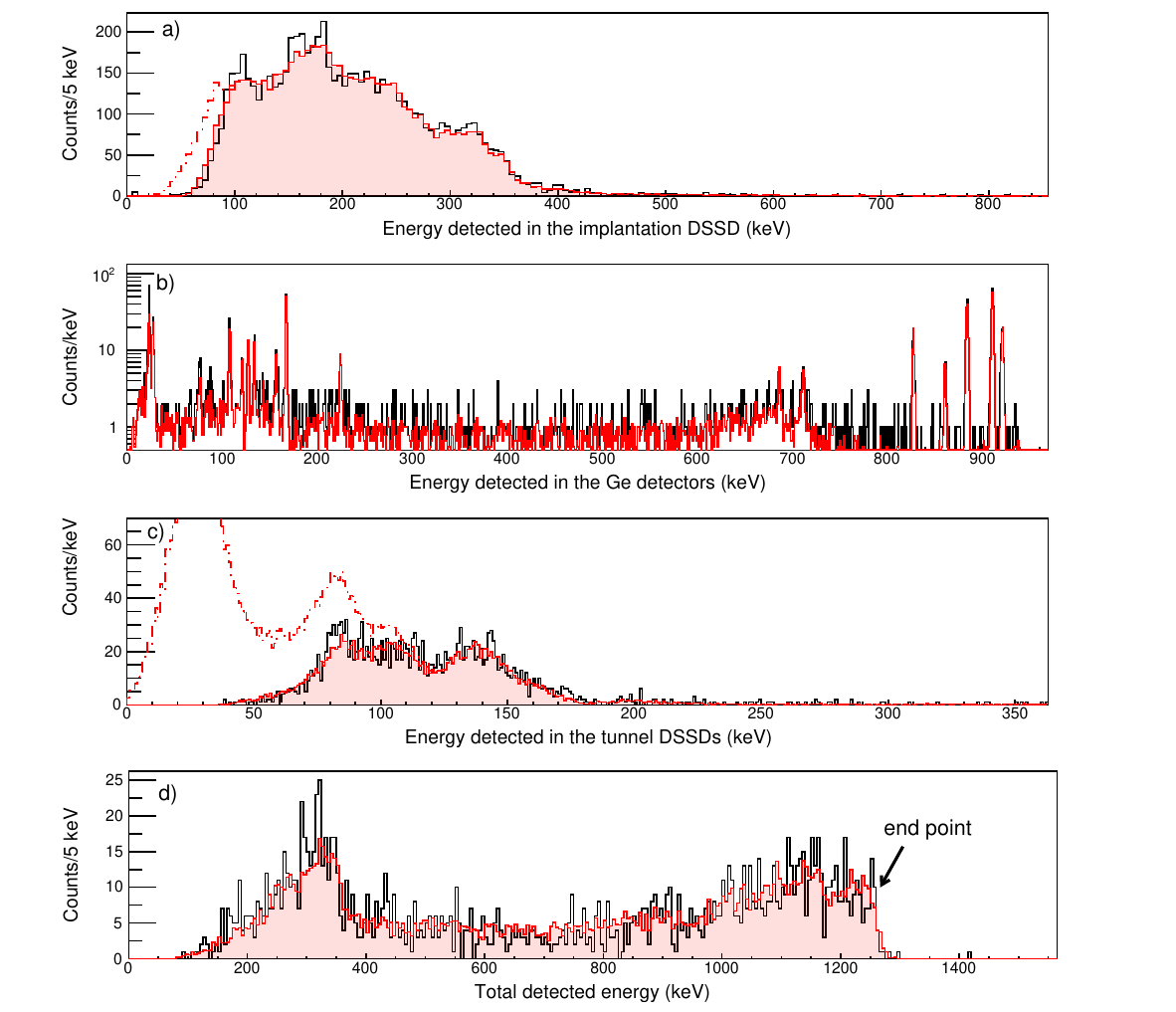} 
   \caption{Simulated (red) and experimental (black) spectra of the decay of $^{252m}$No. a) Energy spectrum of [F] and [FB] events detected in the implantation DSSD, b) spectrum of photon energies detected in the Ge detectors in coincidence with implantation DSSD events, c) spectrum of ICE energies observed in the tunnel DSSDs in coincidence with implantation DSSD events and d) sum of all the energies detected in the GABRIELA detector array when a photon is detected. The dashed red lines are the raw DSSD simulated spectra before the application of the corresponding sigmoid-like thresholds.}
   \label{sim252No}
\end{figure}

Figure \ref{sim252No} shows the comparison between simulated and experimental spectra of detected energies in all the different detectors of GABRIELA in the case of the decay of $^{252m}$No decay. A unique normalisation factor has been applied to all the simulated spectra: the ratio between the simulated and experimental number of detected events in the implantation detector. The global features of all the spectra are qualitatively well reproduced and in particular, the sigmoid functions used to simulate the effect of thresholds in the tunnel and implantation DSSDs properly describe the data at the lowest energies. Moreover, the simulated spectrum of the total energy detected in coincidence with $\gamma$ rays ends at the same point as the experimental one and the end point corresponds, as it should, to the measured excitation energy of the 8$^-$ isomer: 1255 keV.

\begin{figure}[htbp]
   \centering
  \includegraphics [angle=0,width=0.5\textwidth]{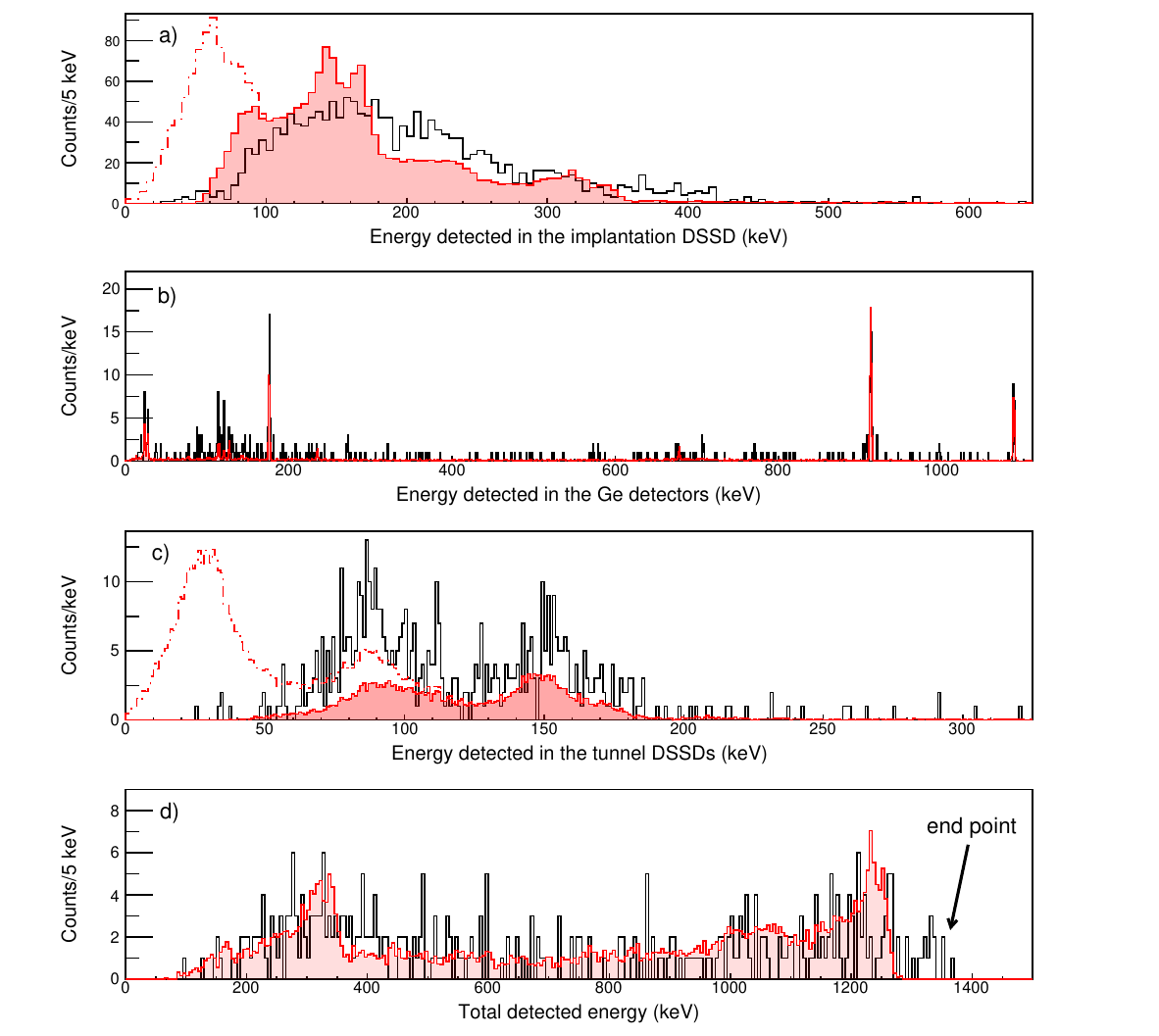} 
    \caption{Same as in figure \ref{sim252No} for $^{250m}$No.}
   \label{sim250No}
\end{figure}

In order to compare with the situation in $^{252m}$No, the same experimental spectra were produced for $^{250m}$No and are compared to their simulated counterparts in figure \ref{sim250No}. It is to be noted that given the 4-8 $\mu$s dead time and the timing condition imposed on the isomeric DSSD signal with respect to the recoil implantation, many $\gamma$ rays and ICEs emitted in the decay of $^{250m}$No are detected in the absence of a [F] or [FB] signal in the DSSD. Requiring a coincident implantation detector signal, as was done for $^{252}$No, leads to a reduction of the statistics in the Ge and tunnel DSSD spectra compared to those of figure \ref{fig5}. The normalisation between simulated and experimental spectra was performed in a similar way as for $^{252}$No. However, the Ge and ICE simulated spectra had to be further multiplied by a common factor of 60$\%$ to bring the simulated intensity in the 912 and 1088~keV peaks to the level of the experimental intensities. This mismatch is an indication that the decay scheme of $^{250m}$No in figure \ref{schemes} is not correct. Furthermore, it is clear from the top and bottom panels of figure \ref{sim250No} that there is a lack of deposited energy in the implantation DSSD and that the end point of the simulated total energy spectrum is underestimated by $\approx$85-90 keV. This means that the $\approx$40 $\mu$s isomer lies above the 6$^+$ state in the bottom panel of figure \ref{schemes}.

\begin{figure}[htbp]
   \centering
  \includegraphics [angle=0,width=0.5\textwidth]{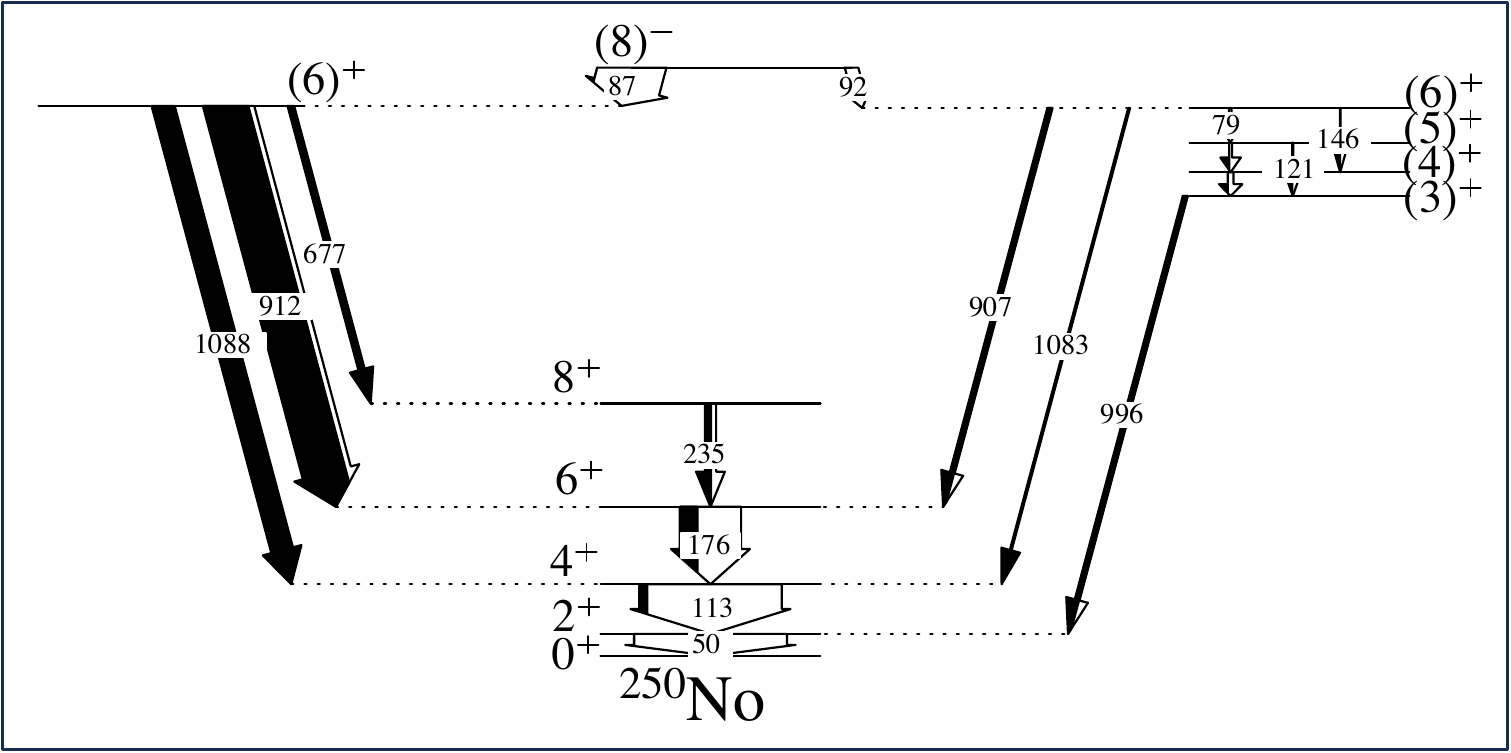} 
  
   \caption{Proposed partial decay scheme of $^{250m}$No. See text for details}
   \label{schemerev}
\end{figure}

Since a 2 quasi-proton 3$^+$ state is observed at an excitation energy of $\sim$1 MeV in $^{254}$No \cite{Herzberg2,Tandel,Clark10,Hessberger2,Wahid}, it is likely that such a state also lies at low energy in $^{250}$No since it has the same number of protons as $^{254}$No. Similarly, one expects the presence of a low-lying 8$^-$ state, which is present in the heavier No isotopes. In $^{254}$No, the nature of the 8$^-$ isomer at 1297 keV is still under debate while for the 8$^-$ state at 1255 keV in $^{252}$No, its two-quasi-neutron character based on the 7/2$^+$[624] and 9/2$^+$[734] Nilsson states has been established via prompt $\gamma$-ray spectroscopy \cite{Sulignano2}. Taking the 996 keV transition to be the transition connecting the 3$^+$ band head to the 2$^+$ member of the ground state rotational band in $^{250}$No, the $K$=3 band is extended to higher spin using a moment of inertia obtained by scaling the moment of inertia of the 3$^+$ band in $^{254}$No by a factor 
1.15 (a factor only slightly larger than the ratio of ground-state-band moments of inertia). In this scenario, the 6$^+$ member of the 3$^+$ rotational band would to lie 5 keV below the 6$^+$ state of figure \ref{schemes}. Consequently, the weak 907 and 1083~keV transitions could be interpreted as de-exciting the lower-lying 6$^+$ state, mirroring the 912 and 1087~keV decays from the higher-lying 6$^+$ state. $K$ mixing between the $K$=6 and $K$=3, 6$^+$ states could also account for the fact that the decay from the higher-lying 6$^+$ state to the ground-state rotational band appears as prompt within our timing resolution. Indeed, a $\Delta K$=6 pure M1 transition of 912~keV would have a half life of $\approx$300 $\mu$s while a $\Delta K$=3 transition of similar energy would be 10$^6$ times faster according to the Lobner systematics \cite{Lobner}. Admixtures of the $K$=3, 6$^+$ state larger than 10$^{-4}$ would therefore result in sub-$\mu$s half lives in the case of a pure M1 transition. In the new scenario, both 6$^+$ states are  populated by the decay of the isomer which lies 85-90 keV above them and which we assume to be an 8$^-$ state. The new decay scheme proposed for $^{250m}$No is shown in figure \ref{schemerev}. The 8$^-$ isomer has an excitation energy of 1338 keV and decays by M2 transitions to the 6$^+$ states below it. An 87~keV M2 transition is expected to have a lifetime of the order of 0.4 $\mu$s while the isomer has a half life of $\approx$40 $\mu$s. The retardation factor of 100 can be attributed to $K$ mixing and the microscopic rearrangement required to make the transition from the two-quasiparticle 8$^-$ state to the 6$^+$ state. 

\begin{figure}[htbp]
   \centering
  \includegraphics [angle=0,width=0.5\textwidth]{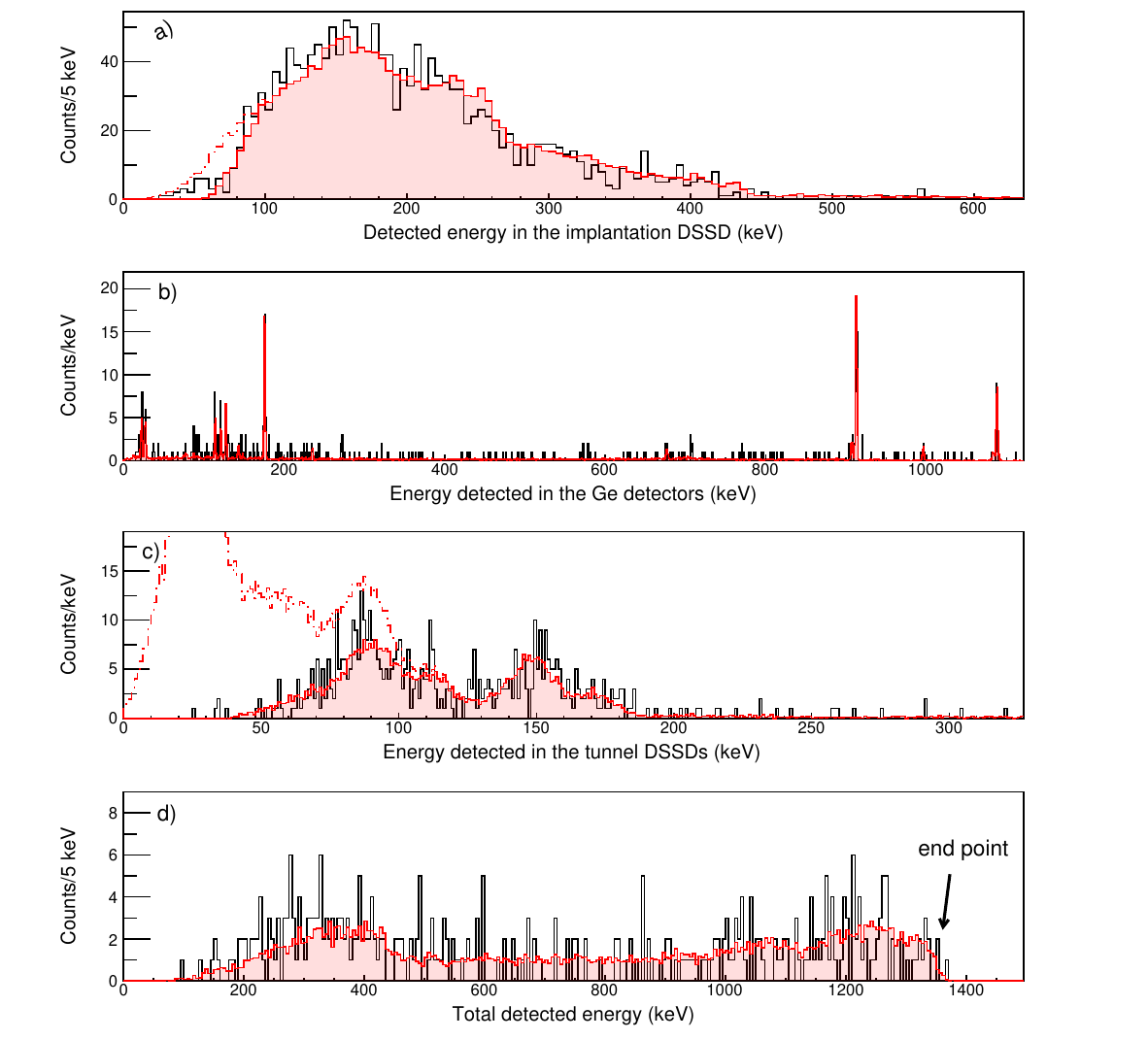} 
    \caption{Same as in figure \ref{sim252No} for $^{250m}$No using the revised decay scheme of Fig \ref{schemerev}.}
   \label{sim250Nonew}
\end{figure}

The decay scheme of figure \ref{schemerev} has been put through the Monte Carlo simulation test. The comparisons between simulated and experimental spectra are shown in figure \ref{sim250Nonew}.
The main characteristics of the experimental spectra are well accounted for with a unique normalisation factor, giving confidence in the proposed decay scenario. Assuming that the population of the 6$^+$ states is governed by the $K$=6 content in their wave functions, an 86$\%$ $K$=6 admixture can be inferred for the 6$^+$state emitting the 912 and 1088 keV $\gamma$ rays and a half life of $\approx$2 ns can be deduced from the corresponding $K$=3 admixture. Such a half life is beyond our timing resolution and it is also beyond what can be resolved in the waveforms of refs. \cite{Kallunkathariyil,Batar250No}.
The 707 and 131 keV transitions visible in the $\gamma$-ray spectrum of figure \ref{fig5} could not be placed in the revised scheme of figure \ref{schemerev}. The peaks lie in regions of the spectrum where the density of background counts is larger due to the Compton edge and backscatter peak (see panel b) of figure \ref{sim252No}). They could therefore correspond to statistical fluctuations. If, on the other hand, the peaks are real, this may indicate that another weaker path is involved in the decay of $^{250m}$No, possibly involving the 2$^-$ phonon band present below 1 MeV in the N=150 isotones \cite{Robinson} and which should also be low-lying in $^{250}$No.

More data is needed to unambiguously establish the decay scheme of $^{250m}$No, but the observed decay properties, especially the end-point of the total-energy spectrum and the shape of the spectrum of deposited energies in the implantation detector, are not compatible with a direct decay to the ground state rotational band. The decay clearly samples intermediate excited states before reaching the ground state. A similar situation could well occur in $^{252}$Rf. The three signals associated with the internal decay of the 13 $\mu$s isomer in $^{252}$Rf have energies equal to or greater than 220 keV (the reported energies in ref. \cite{Batar252Rf} are 0.22, 0.28 and 0.32 MeV). As can be seen from the dashed red spectrum in the top panel of figure \ref{sim250No}, the bulk of the spectrum of deposited energies in the DSSD lies below the sum of the ground-state band  2$^+-0^+$ and 4$^+-2^+$ transition energies. A similar sum energy is expected in the case of $^{252}$Rf (see figure \ref{bands}). From the fraction of events lying above 220 keV in the dashed spectrum, it can be deduced that the probability of a direct decay to the ground state band yielding a signal greater than 220 is 15$\%$ (depending on the implantation depth, these numbers may vary a little, but they will remain small). The probability of detecting three such events is therefore not likely and could already be an indication that the decay of the 13 $\mu$s isomer in $^{252}$Rf is indeed not direct. Moreover, $K$ mixing between the intermediate states sampled in the internal decay of the isomer could lead to internal-decay half lives ranging from sub-ns to several $\mu$s depending on the $K$-quantum numbers of the states below the isomer and the details of the $K$ mixing at play. The fact that a 60 ns fission signal is observed after the 13 $\mu$s internal decay could therefore be interpreted as evidence for a 60-ns-delayed ultra-fast fission of the ground state, in line with what was concluded in ref. \cite{LopezRf}. 
\section{Conclusion and perspectives}

A \textsc{Geant4}-assisted analysis of the internal decay of $^{250m}$No has revealed that the isomer lies at higher excitation energy than what was previously thought and that it does not decay directly to the ground state rotational band. A new partial decay scheme has been proposed, which reproduces all the spectroscopic features of the decay. In this new decay scenario, the isomer lies at an excitation energy of 1338 keV and is assigned a $K^\pi$=8$^-$ configuration. Its internal decay most likely samples $K$-mixed intermediate states before reaching the ground state. Experimental evidence has been put forward suggesting that such a complex isomeric decay might also occur in the isotone $^{252}$Rf. This implies that the assignment of the 60~ns fission activity to the ground state activity of $^{252}$Rf remains speculative and that further investigations are required to definitely establish the limit of stability of neutron-rich Rf isotopes.

\section*{Acknowledgements}

This work was supported by the Russian Foundation for Basic Research (project no. 17-02-00867 and 18-52-15004), the French national Research Agency (projects nos. ANR-06-BLAN-0034-01 and ANR-12-BS05-0013) and the IN2P3-JINR collaboration agreement no. 04-63.


\begin{thebibliography}{}
\bibitem{Hessberger17} F.P. Hessberger et al., Eur. Phys. J. A 53, 75 (2017) 

\bibitem{Xu} F. Xu et al. Phys. Rev. C 92, 252401 (2004) 
\bibitem{Jachimowicz} P. Jachimowicz, M. Kowal, and J. Skalski Phys. Rev. C 92, 044306 (2015) 
\bibitem{Clark} R.M. Clark, EPJ Web of Conferences 131 (2016) 02002
\bibitem{Kondev} F. Kondev, G. Dracoulis and T. Kibedi, At. Dat. And Nucl. Dat. Tables 103-104 (2015) 50
\bibitem{David} H. David et al., Phys. Rev. Lett. 115, 132502 (2015) 
\bibitem{Kallunkathariyil} J. Kallunkathariyil et al. Phys. Rev. C 101, 011301(R)  (2020) 
\bibitem{Batar250No} J. Khuyagbaatar et  al., Phys. Rev. C 106, 024309 (2022)  
\bibitem{Batar252Rf} J. Khuyagbaatar et al., Phys. Rev. Lett. 134, 022501 (2025)
\bibitem{LopezRf}
A. Lopez-Martens et al., Phys. Rev. C 105, L021306 (2022) 
\bibitem{Batar253Rf} J. Khuyagbaatar et al., Phys. Rev. C 104, L031303 (2021)
\bibitem{Rogov1} I.S. Rogov, G.G. Adamian, N.V. Antonenko, Eur. Phys. J. A 60, 164 (2024) 
\bibitem{Rogov2} I.S. Rogov, G.G. Adamian, N.V. Antonenko, Phys. Rev. C 110, 014606 (2024)
\bibitem{Moller}
P. M\"oller, J.R. Nix, W.J. Swiatecki, Nucl. Phys. A 469 (1987) 1
\bibitem{Rodriguez} 
R. Rodriguez-Guzm\'an and L.M. Robledo, Eur. Phys. J. A 52, 348 (2016) 
\bibitem{Pu} E.M. Franz et al., Phys. Rev. C 23, 2234 (1981)
\bibitem{Cm} P.G. Hansen et al., Nucl. Phys. 45 (1963) 410

\bibitem{Tezekbayeva} M. S. Tezekbayeva et al., Eurasian J. Phys. Func. Mat. 3, 300 (2019)
\bibitem{Smolanczuk}R. Smolanczuk, J. Skalski, and A. Sobiczewski, Phys. Rev. C 52, 1871 (1995)
\bibitem{Batar257Sg} P. Mosat et al., Phys. Rev. Lett. 134, 232501 (2025)
\bibitem{Belozerov} A. Belozerov et al., Eur. Phys. J. A 16, 447 (2003) 
\bibitem{Peterson} D. Peterson et al., Phys. Rev. C 74, 014316 (2006) 
\bibitem{Svirikhin} A. Svirikhin et al., Physics of Particles and Nuclei Letters, Vol. 14 (2017) 571
\bibitem{geant4}A. Agostinelli et al. Nucl. Instr. Meth.  A 506 (2003) 250

\bibitem{Hauschild} K. Hauschild et al., Nucl. Instr. Meth. A 560 (2006) 388
\bibitem{Chakma} R. Chakma et al., Eur. Phys. J. A 56, 245 (2020)  
\bibitem{LopezGabriela} A. Lopez-Martens and K. Hauschild, Eur. Phys. J. A 58, 134 (2022)

\bibitem{Sulignano} B. Sulignano et al., Eur. Phys. J. A 33, 327 (2007) 
\bibitem{Sulignano2} B. Sulignano et al., Phys. Rev. C 86, 044318 (2012) 
\bibitem{Popeko}A. Popeko et al., Nucl. Instr. Meth. B 376  (2016) 140 
\bibitem{Rezanka} I. Rezanka, I.M. Ladenbauer-Bellis, J.O. Rasmussen, Phys. Rev. C 10, 766 (1974) 
\bibitem{NNDC}
ENSDF: https://www.nndc.bnl.gov/ensdf/
\bibitem{BRICC} T. Kib\'edi, T.W. Burrows, M.B. Trzhaskovskaya, P.M. Davidson, C.W  Nestor Jr., Nucl. Instr. Meth. A 589 (2008) 202-229
\bibitem{Liu} H.L. Liu et al., Phys. Rev. C 89, 044304 (2014) 
\bibitem{Chakma25}R. Chakma et al., Nucl. Instr. Meth. A 1072, 170144 (2025)
\bibitem{Reiter} P. Reiter et al., Phys. Rev. Lett. 82, 509 (1999)
\bibitem{Leino} M. Leino et al., Eur. Phys. J. A 6, 63 (1999)
\bibitem{Butler} P.A. Butler et al., Phys. Rev. Lett. 89, 202501 (2002)
\bibitem{Eeckhaudt} S. Eeckhaudt et al., Eur. Phys. J. A 26, 227 (2005) 
\bibitem{Herzberg} R-.D. Herzber et al., Phys. Rev. C 65, 014303 (2001) 

\bibitem{Harris} S. Harris, Phys. Rev. 138, B509 (1964)
\bibitem{Raeder}  S. Raeder et al., Phys. Rev. Lett. 120, 232503 (2018)
\bibitem{Greenlees} P.T. Greenlees et al., Phys. Rev. C 78, 021303(R) (2008)  
\bibitem{Ketelhut} S. Ketelhut, PhD thesis, University of Jyv\"askyl\"a (2010)
\bibitem{Piot} J. Piot et al., Phys. Rev. C 85, 041301(R) (2012)
\bibitem{GreenleesRf} P.T. Greenlees et al., Phys. Rev. Lett. 109, 012501 (2012)
\bibitem{Seweryniak} D. Seweryniak et al., Phys. Rev. C 107, L061302 (2023)
\bibitem{Herzberg2} R.-D. Herzberg et al., Nature 442 (2006) 896
\bibitem{Tandel} S.K. Tandel et al., Phys. Rev. Lett. 97, 082502 (2006) 
\bibitem{Clark10} R.M. Clark et al., Phys. Lett. B690 (2010) 19-24
\bibitem{Hessberger2} F.P. Hessberger et al., Eur. Phys. J. A 43, 55 (2010) 
\bibitem{Wahid} S. G. Wahid et al., Phys. Rev. C 111, 034320 (2025)

\bibitem{Lobner} K.E.G. L\"obner, Phys. Lett. B 26 (1968) 369
\bibitem{Robinson} A. P. Robinson et al., Phys. Rev. C 78, 034308 (2008) 



\end{thebibliography}
\end{document}